\documentclass[oneside]{article}
\usepackage[T1]{fontenc}
\usepackage{authblk}
\usepackage{float}
\usepackage{amsmath}
\usepackage{graphicx}
\usepackage{xfrac}
\usepackage[english]{babel}
\usepackage{lmodern}
\usepackage[T1]{fontenc}
\usepackage[latin9]{inputenc}
\usepackage{mathtools}
\usepackage{graphicx}
\usepackage{esint}
\usepackage[unicode=true,
 bookmarks=false,
breaklinks=false,pdfborder={0 0 1},backref=false,colorlinks=false]
 {hyperref}
\usepackage[a4paper, total={210mm,297mm},margin=2.0cm]{geometry}

\usepackage{datetime}
\newdate{date}{23}{11}{2018}

\title{Of Naturalness and Complexity}

\author[1,2,3]{Sauro Succi \thanks{Electronic address: \texttt{sauro.succi@gmail.com}; Corresponding author}}

\affil[1]{Center for Life Nanoscience at La Sapienza, Fondazione Istituto Italiano di Tecnologia, Roma, Italy}
\affil[2]{Institute for Applied Computational Science, J. Paulson School of Engineering and Applied Sciences, Harvard University, Cambridge USA}
\affil[3]{Istituto per le Applicazioni del Calcolo CNR, Via dei Taurini 19, 00185 Rome, Italy}

\date{\displaydate{date}}

\begin{document}

\maketitle

\begin{center}
{\it Elegance is the only beauty that never fades (A. Hepburn)}
\end{center}

\begin{abstract}
It is argued that the occurrence of disproportionately ("un-natural") large (or small) numbers, as well as 
deep cancellations, are comparatively natural traits of the way Nature is geared to operate 
in most complex systems.
The idea is illustrated by means of two outstanding and over-resilient problems in theoretical physics:
fluid turbulence and the computation of ground-states of quantum many-body fermion systems.
Potential connections with the issue of Naturalness, or lack thereof, in high-energy physics are sketched out.
\end{abstract}
 
\section{Introduction}

In a recent paper, Giudice discusses the prospects of the post-naturalness era, namely
the future of physics beyond the standard model, in case the guiding principle of Naturalness 
goes unattended by post-Higgs experimental searches \cite{GG1}.
Among other inspiring items, the author points out that periods of Krisis (K not C, standing for opportunities 
before it does for problems) offer a fertile ground for cross-contributions from different areas of physics,
in fact, possibly even beyond the realm of physics.

What follows is a modest tentative precisely along this line, from the outside of the hep community.

In particular, we wish comment on the fact that lack of Naturalness, meaning by this sensitivity
of large-scale parameters to the small-scale ones, may just  be a comparatively "natural" consequence 
of some typical hallmarks of complexity, particularly Non-linearity, Hyper-Dimensions and Non-locality,
three infamous barriers to human understanding altogether. 

We also comment on Sneakiness, namely the tendency of complex systems to segregate 
key processes within very small, sometimes astronomically small, regions of their 
phase space, as a consequence of three barriers above.

Connections between Sneakiness and Fine-Tuning are also drawn.

It is also observed that the hallmarks of Complexity, such as inter-scale sensitivity, sneakiness and rare events, 
often act as effective generators of highly un-natural numbers, which may eventually near-cancel due 
to a tight balance between the competing mechanisms.  
Since it is precisely this tight balance which fuels complexity, deep cancellations, rather 
than standing as a "ugly" anomaly, might speak instead for the mechanisms by which nature "plays hide-and-seek" 
with her deepest and subtlest phenomena \cite{NAT}.

\section{The "problem" of Naturalness}

Naturalness is usually referred as to the "Principle" according to which effective parameters of a low
energy (effective) field theory should prove largely insensitive to the actual details and 
values of the parameters of the underlying microscopic (high-energy) theory.
In broad strokes, a close relative is known as Universality in the statmech community \cite{OUP2018}.

The notion of Naturalness connects to a variety of leading avenues of theoretical hep, such as 
Supersymmetry (SUSY), Renormalisation Group, Symmetry, Fine-Tuning,  up to the philosophically-bent
digressions on the alleged Beauty\&Truth equivalence, to name but a few among the most debated ones.
The problem of Naturalness has been around for a long time in high-energy physics, but it has become 
particularly acute in the last few years, mostly on account of the missing evidence of SUSY 
s-particles in post-Higgs experimental search. So far, at least.

In a recent paper, P. Williams argues that possibly the most productive way of thinking of 
Naturalness is to discuss its connections with the issue of Interscale Sensitivity (IS) \cite{PW}. 
This ties in with Giudice's remark that failure of Naturalness may open up a new
post-naturalness era, in which central stage should be taken by correlations
that manage to persist across widely separated scales.
Once again, this evokes an over-arching theme of theoretical physics besides and beyond hep, 
namely the issue of scale-separation and the steep problems that arise whenever
scales show no wish to separate, fluid turbulence being a prime example in point. 

In this paper, we wish to raise the point that other areas of theoretical physical, most notably 
the science of complex (natural) systems, being inherently confronted with the problem of 
lack of scale separation, may bring some useful ideas and tools to the fore of post-naturalness physics.

In particular, we shall argue that once the notion of  {\it non-linear cascade} 
as a natural scale bridger, is imported within the hep framework, some cases of
missing Naturalness may appear less surprising than they currently seem.
Likewise, the tendency of hyper dimensional systems to generate highly 
localised (sneaky) distributions, may also give rise to the occurrence of  "un-natural" numbers. 
Finally, we shall argue that deep cancellations are not unique to hep or cosmology, the ground state
of many-body fermions offering an adamant example in point, with cancellations which may readily 
exceed the 120 digits of the cosmological constant. 

Having clarified the context, let us begin by reviewing the basic notions of the cradle of turbulence, namely hydrodynamics.

\section{Hydrodynamics}

By name, hydrodynamics should preoccupy itself only with the motion of water. 
To the contrary, hydrodynamics provides the basic theory for fluids in general, water, air, oil and, with due 
precautions, blood, honey, and ketchup as well.
In a word, hydrodynamics is {\it universal},  i.e. the structure of its basic equations is 
independent of the details of the underlying molecular interactions.
The basic equations of fluid-dynamics are no spring-chicken, written as they were no less
than two centuries ago by the anglo-irish physicist-mathematician
Gabriel Stokes (1819-1903) and the french physicist-engineer Claude-Louis Navier (1785-1836). 
Ever since, they have not ceased to provide uncountable insights into the study of fluid motion across
an amazingly broad spectrum of applications (don't forget, fluids are everywhere, outside and within us...). 

Here come the Navier-Stokes equations in their full splendor:
\begin{eqnarray}
\label{NSE}
\partial_t \rho + \partial_i (\rho u_i) = 0\\
\partial_t (\rho u_i) + \partial_j P_{ij} = f_i
\end{eqnarray}
where latin indices run over spatial dimensions and repeated indices imply summation.

In the above, $\rho$ is the fluid density, $u_i$ the flow velocity, $P_{ij}$ is pressure tensor
and $f_i$ is an external force (per unit volume).
The first equation is just a mass conservation statement, while the second is basically
Netwon equation in reverse $m a_i=F_i$, both applied to a finite volume of fluid. 

The pressure tensor counts three contributions
$$
P_{ij} = \rho u_i u_j - p \delta_{ij} + \sigma_{ij}
$$
The first term at the right hand side is the fluid inertia, whose divergence gives i.e the acceleration of the
fluid element along the Lagrangian trajectory defined by the fluid velocity itself (self-advection).
The second term is the fluid pressure and the third is the stress tensor, encoding dissipative effects.
For many fluids of interest, this latter can be expressed as a linear combination of the velocity 
gradients (so-called Newtonian fluids): 
\begin{equation}
\label{STRESS}
\sigma_{ij} = 2 \mu (\partial_i u_j + \partial_j u_i) + \lambda (\partial_k u_k) \delta_{ij} 
\end{equation}
where the coefficient $\lambda$ and $\mu$ associate with the shear and bulk viscosity of the fluid.
The constitutive relation eq. (\ref{STRESS}) is the -only- bridge between the macroscopic fluid and the
underlying molecular world.

Note that in the inviscid limit, the fluid equations (then known as Euler equations) reflect 
in full the symmetries of Newtonian mechanics.
In particular, they obey PT invariance, i.e. space and time inversion ($x \to -x$, $t \to -t$, $u_i \to u_i$) 
as well as scale invariance, ($x \to k^a x$, $t \to k^b t$, $u \to k^{a-b} u$).

Both symmetries are broken by the dissipative term, the reason being that such term does {\it not}
descend from reversible Newtonian mechanics, but represents a weak-gradient approximation instead, 
whereby stress and the resulting strain are assumed to be linearly proportional
\footnote{ In hindsight, it is rather ironic that fluids obeying this linear relation are called Newtonian, in the face of the
fact that the dissipative term is precisely the only one which cannot be derived from Newtonian mechanics!}.

Despite their conceptual simplicity, the Navier-Stokes equations continue to pose a 
formidable mathematical challenge (worth a million dollar Clay Prize), the main culprit
being {\it Turbulence}, the name of hydrodynamics whenever inertia overwhelms dissipation.  

In the last decade, hydrodynamics has received a major boost of popularity outside the fluid community, mostly 
on account of the celebrated AdS-CFT duality, which states the equivalence between 
$(d+1)$-dimensional gravity and $d$ dimensional field theories, most notably hydrodynamics \cite{MALDA}.
More specifically, it has been realised that strongly interacting quantum-relativistic matter abides
by hydrodynamic behaviour (collective motion) over an amazingly broad range of conditions, from
ultra-dense, ultra-hot quark-gluon plasmas remnant of the early Universe, all the way to ultra-cold Bose 
Einstein condensates, spanning nearly the entire density-temperature spectrum of the current Universe.
An unprecedented triumph of Universality!
 
But this another paper, so let us now move on to turbulence. 

\subsection{Fluid Turbulence}

Turbulence is the state of flowing matter characterised by the dominance of Inertia versus Dissipation.
This results in a progressive loss of coherence in time first, and later on, space as well, which 
is a byname for poor predictability, as best epitomised by weather forecast
(to go with Dylan, "the answer, my friend, is blowing in the wind"..). 
Note, turbulence is neither chaos, nor blind randomness, but rather a subtly organised 
network of non-linear and non-local (classical) correlations between events taking place over a 
broad range of scales in space and time \cite{TURBO}.
The simplest instance of  homogeneous (no boundaries) incompressible (no divergence) flow, turbulence
is governed by a single dimensionless parameter, known as Reynolds number: 

\begin{equation}
\label{REY}
Re = \frac{UL}{\nu}
\end{equation}

where $U$ is a macroscopic velocity of a body of macro size $L$ and $\nu$ is the kinematic 
viscosity of the fluid, a genuinely molecular property.

A simple dimensional argument shows that Reynolds measures the relative strength between non-linearity,
the kinematic acceleration, $u_j \nabla_j u_i$ versus viscous dissipation $\nu \nabla^2 u_i$. 

Turbulence is the name of hydrodynamics  in the limit of very large Reynolds numbers.

But, how large is large, here?

The calculation is simple: take an ordinary car, say $2m$ long, at a standard
speed of $30$ m/s: with kinematic viscosity of air at $\nu \sim 1.5 \;10^{-5}$, 
the above formula gives  $Re \sim  \frac{30 \times 2}{ 1.5 \;10^{-5}} = 4 \; 10^6$.   

Now let's turn to field theory, where a glorious name for hydrodynamics is  
quadratic self-interacting, non-local, classical vector field theory in $d+1$ dimensions.
The Reynolds number is the coupling strength of the flow field with itself, since 
the fluid moves momentum along material lines defined by the fluid momentum itself.
While the QED analogue is  $1/137$ and QCD about $1$, as we have just seen, 
turbulence typically features into the millions even for ordinary objects, such as a "humble" car.
Geophysics goes easily orders above, and going off the planet, out there in the Universe,
astrophysical and cosmological flows take us much farther than that.

No surprise that two centuries down the line, the problem is still with us.

Next, let us ask a naive question: {\it why} is the Reynolds number so "un-naturally" large?

Little elaboration of the expression (\ref{REY}) unveils the point.
By plain dimensional arguments, we write $\nu \sim l_m \bar v$, 
where $l_m$ is the molecular mean free path and $\bar v$ is the thermal speed of the 
underlying molecules, thus yielding  $Re \sim (U/\bar v) \times (L/l_m)$. 
Now take $\bar v$ of the order of speed of sound (true for ideal gas
and not far-off for most fluids), and what we get is a nice duality 
\begin{equation}
\label{KARMAN}
Re =Ma/Kn
\end{equation} 
also known as von-Karman relation, where $Ma=U/c_s$ is the Mach number, the
ratio of fluid to sound speed and $Kn=l_m/L$ is the Knudsen number.
The latter measures the scale separation between the micro and macroscopic world and it
is commonly assumed that fluid behaviour starts roughly below $Kn \sim 0.01$, two 
decades above the microscale \footnote{
This makes sense, two decades in three dimensions imply one million molecules per
fluid degree of freedom, hence roughly a $10^{-3}$ statistical uncertainty.  
}.
 
Next, less foregone than it seems, {\it macroscopic bodies can move as fast as molecules}, 
that is precisely what a supersonic airplane does!  
Thus, speed is no divide between the micro and the macro worlds.
For an ordinary car, Mach is the order of $0.1$, thus meaning that Reynolds is 
still just one tenth of the inverse Knudsen
\footnote{We discount the Mach number, by taking it $O(1)$, which is not that wrong for many 
macroscopic flows, where $Ma \sim 0.1$.}.
 
Hence, Reynolds is large because it measures the separation between the
macro and micro levels; in actual facts, it measures the size of our 
car in units of the molecular mean free path!
For air in standard conditions $l_m \sim 50$ nm, which immediately delivers 
seven orders of magnitude, hence Reynolds in the millions at $Ma \sim 0.1$. 

These units are utterly "un-natural"; to be sure, they look decidedly awkward, 
it is as if we were measuring our height in units of the molecules we are made of (the result would
be of the order of one billion, not much above the Reynolds number of airplanes or ships). 
Yet, these are nonetheless the right units to capture the competition between inertia 
and dissipation, i.e. the very roots of the complexity of fluid turbulence. 

The key point is that Inertia is an exquisitely macroscopic mechanism 
(in the micro and nano worlds matter moves by entirely different mechanisms), while Dissipation 
is strictly controlled by molecular properties, which dictate the value of the kinematic viscosity.

Yet, they are two inseparable faces of the same medal, which is, in the end, the
ultimate root of the spectacular complexity of turbulent flows.

So, let us dig a bit deeper into these matters.

\subsection{Coherence length and the Kolmogorov cascade}

Turbulence is a quintessential expression of the "IR-UV" coupling between Inertia and Dissipation, the competition
between these two grand-contenders resulting in a
broad and {\it gapless} spectrum of excitations, running from the macro-scale $L$, all the way down to the dissipative 
scale $l_k$, also known as Kolmogorov length, after the russian polymath Andrei Kolmogorov (1903-1987).

Here, a basic question arises: what is the basic mechanism in control of such dissipative length, namely the smallest 
lengthscale compatible with the coherent structures (eddies) which support turbulence?
Naively, one could identify such length with the mean-free path, as we just did when asserting that
Reynolds measures our car in units of the mean free path.
This is approximately true on numerical grounds, but conceptually very incorrect, because
there is no collective motion at the scale of the mean-free-path!

By definition the Kolmogorov length is precisely the scale 
at which inertia and dissipation come to an exact balance:
\begin{equation}
Re(l_k) =1
\end{equation}
In passing, we note that as in any field-theory,  the Reynolds number is a running coupling constant. 
How does one extract $l_k$ from the above definition?

To this purpose, one needs to know the value of the velocity field at the Kolmogorov scale, which 
is tantamount to solving the equations of motion! 

Kolmogorov took a brilliant shortcut instead: by playing a simple argument of scale invariance 
of the energy {\it flux},  $u^3(l)/l$ (the rate of kinetic energy transfer), Kolmogorov came up 
with a utterly simple expression for the dissipative length, namely
\begin{equation}
\label{G34}
l_k = L/Re^{3/4}
\end{equation}

A direct comparison with (\ref{KARMAN}), shows that  

\begin{equation}
\label{G14}
l_k/l_m\sim Re^{1/4}
\end{equation} 

thus informing us that the separation between the smallest collective scale and the 
mean free path grows at increasing Reynolds.

But the separation between the largest and smallest macroscales grows much 
faster, namely:  
\begin{equation}
\label{G3}
\frac{L}{l_k} = (\frac{l_k}{l_m})^3
\end{equation}
In other words, the coherence length is much closer to the mean free 
path than it is to the macroscale $L$. This is why continuum mechanics can  
extend down to pretty small supra-molecular scales.

At $Re=10^7$, a very fast car, the spectrum covers five decades, say from
1 meter down to some 10 microns, before dissipation takes over in the two-three underlying decades.

This clearly informs us that the physics of turbulence and, for that matter, of any
strongly interacting field theory,  "cannot be examined shell-by-shell" \cite{GG1}, the
reason being that the spectrum is not only broad but, most importantly, it is {\it gapless}!
Eddies interact preferentially with eddies close in size, locality in $k$-space, but the 
effects of small eddies on the large ones cannot be ignored because they 
propagate across a continuum spectrum, the celebrated {\it turbulent cascade}.

Summarising, we have three basic scales: the macro scale $L$, at which energy is injected (IR),
the microscale $l_m$ associated to the microscopic degrees of freedom (UV) and  an
intermediate, "mesoscopic" scale, (Kolmogorov) $l_k$, which lies in-between.

In the sequel, it shall prove expedient to regard the Kolmogorov scale from a broader perspective, 
i.e. as a {\it coherence} scale $l_c$,  below which the nonlinear energy cascade is no longer able 
to sustain coherent (collective) motion.
In a nutshell, coherence lives in the range $L>l>l_c $, and what is left below $l_c$ is "chaotic" motion,
i.e. energy is irreversibly dissipated, never to come back in organised patterns.
It is only "natural" (no pun), to interpret $l_c$ as the smallest scales capable of 
showing correlation with the macroscale $L$.
Note that the coherence length also fixes the lower bound for continuum field theory 
to apply in general, not only turbulence.

After these preparations, we are finally ready to ask the main question: is Turbulence "Natural"?

If by Naturalness we imply  that the physics of turbulence, which lives by definition 
in the range $l_c < l <L$, should be largely unaffected by the details of the physics in 
the microscale range $l_m < l < l_c$, then, yes, turbulence is natural indeed. 
It is in fact generally accepted that turbulence is a quintessential collective 
phenomenon, oblivious of molecular details (Universality)
\footnote{
With a notable exception though; it cannot be ruled that
details of micro-corrugations on solid walls, may have a long-lasting effect 
on the way turbulence  is excited in the first place and how it develops subsequently.
If confirmed, this would violate Universality.
}.
Of course, molecular details reflect on the Reynolds number, but only through the 
kinematic viscosity, a single number to represent the zillions of molecular complexions underneath!  

On the other hand, if Naturalness means dealing with numbers of order unity and 
close relatives thereof, then turbulence is not natural at all!

As discussed earlier on, the values of the Reynolds number for most common 
flows appear to be un-naturally large, which is by no means a coincidence: it is indeed quite "unnatural" for energy to 
be dissipated many spatial decades below the scale at which it is injected!

But this is precisely  the essence of the glueing power nonlinear cascades: energy cascades from the 
macroscale $L$ down to the dissipative scale $l_k$, at virtually zero-dissipation!
Since dissipation is the only scale-breaking term in the fluid equations, the cascade can
indeed proceed through a scale-invariant energy flux down to $l_c$, which was precisely 
Kolmogorov's intuition.   

Far from being the exception, this interscale dependence is a main hallmark of 
many complex systems driven by strong non-linearities, which brings us to our next 
topic: cascades in field theory. 

\section{Nonlinear cascades in field theory}

Fluid turbulence is an epitome of non-linear cascade, but certainly not the only one.
Any non-linear field theory, classical or quantum, may exhibit its own "turbulence", meaning 
by this loss of coherence both in space and time.

To fix ideas, let us consider a fluctuating scalar $\Phi(x)$ and define
$$
\phi(x;l) \equiv \Phi(x+l) - \Phi(x)
$$
its associated two-point fluctuation.

Next, assume the latter scales like
$$
< |\phi(x;l)| > \sim l^h
$$
where brackets denote some suitable ensemble average, in practice often just space-time averaging.

The scaling exponent $h$ (Hoelder exponent for mathematicians) measures the degree 
of randomness of the fluctuating field. 
The value $h=0$ stands for full randomness, fluctuations are scale-independent, they do not recede 
at decreasing separation, while $h=1$ (fluctuations recede linearly with separation), denotes a 
smooth and differentiable field.
Negative $h$ codes for singularity  (fluctuations increase at decreasing separation), while 
$h>1$ stands for super-smoothness (zero gradient almost everywhere).

Fluid turbulence sits at $h=1/3$, as it follows straight 
from Kolmogorov  assumption of a scale-invariant energy-flux, i.e.  $v^3(l)/l=const$. 
Brownian fluctuations (diffusion) correspond to $h=1/2$, hence turbulence is more irregular than random walks.
The regime $0<h<1/2$ is often referred as sub-diffusive, while $1/2<h<1$ is known as super-diffusive.
In this terminology $h=1$ is often referred as ballistic and $h>1$ as super-ballistic (acceleration). 
 
The power spectrum is the Fourier transform of the variance of 
the fluctuations, hence it scales like 
$$
E(k) \sim k^{-(1+2h)}
$$
The coupling strength $g(l)$ runs with the scale and we set $G \equiv g(L)$, while, in analogy
with fluid turbulence, the  coherence scale is defined by the condition $g(l_c)=1$.

Assuming a running coupling of the form 
$$g(l)=\phi^a(l) l^b,$$  we write $g(l)=(l/l_c)^{\gamma}$, with $\gamma=ah+b$. 

For turbulence $a=1$,$b=1$ and $\gamma=1+h$, $\gamma>0$ standing for UV-free interactions.

Summarizing, the ratio of the macroscopic scale to the coherence length is then given by
\begin{equation}
\label{GAP_MC}
\Lambda \equiv \frac{L}{l_c} = G^{\frac{1}{\gamma}}
\end{equation}
while the complementary ratio of the coherence to the microscopic scale is given by 
\begin{equation}
\label{GAP_Cm}
\lambda \equiv \frac{l_c}{l_m} = G^{\frac{\gamma-1}{\gamma}}
\end{equation}
By construction, $\lambda \Lambda = G$, the full gap, and $\Lambda = \lambda^{\frac{1}{\gamma-1}}$.

As usual, we have discounted non-unit values of the ratio between microscopic and macroscopic speeds.

The borderline case $\gamma=1$ delivers $l_c=l_m$, no micro-macro scale separation. 
This means that nonlinearity can sustain coherence down to the level where 
coherent fluctuations cannot be told apart from the microscopic ones.
For $\gamma<1$, we encounter a paradoxical situation: the coherence length gets {\it smaller} 
than the microscale!  Here, scales are "underlapped", which we take  as a signature of singularity, a signal 
that continuum theory is in need of a microscopic fix (regularisation).

\subsection{Inverse cascades and the hierarchy problem}

The cascades discussed this far are direct, i.e. from large to small scales. 

But the cascade mechanism works both ways, from large to small and viceversa, the latter case
being denoted as {\it inverse} cascade. Instead of large eddies breaking up in small ones, inverse
cascades proceed by merging and coalescence of small eddies into large ones, and are 
usually observed in connection with small-scale energy injection scenarios.  

Inverse cascades are the potentially relevant ones to the hierarchy problem;  an inverse cascade from the Planck length
to an intermediate coherence length between Planck ($\sim 10^{-35}$ m) and Higgs ($\sim 10^{-18}$ m), would 
mitigate the hierarchy problem by a factor $l_c/l_p$.
 
Here, $l_c$ must be regarded with reciprocal spectacles, the largest coherent
length that can be sustained by  aggregation of microscopic degrees of freedom. 
Again, large couplings, combined with $\gamma >1$ exponents, may build-up substantial 
bridges, the fact that SUSY has not (yet) been observed in current ($\sim 10$ TeV) LHC runs, 
rules out just two decades below the Higgs scale, still leaving plenty of room for the 
putative coherence length to settle down well above the Planck scale.  

The fact that two decades already spell fine-tuning for SUSY is, in this 
respect, a rather irrelevant detail, since what really matters is the existence of 
coherence scale substantially above the Planck length, whatever the underlying mechanism is.

The crucial point, though, is that the coupling strength $g(l)$ must contain 
an explicit dependence on the scale $l$. 
No renormalized polynomial on top of the quartic Higgs potential can fill this bill, because ratios 
of polynomials discriminate field amplitudes, not their scale.

The condition $g(l_c)=1$ then defines a coherence scale above the microscale $l_m$ by a factor
$\lambda = G^{\frac{\gamma-1}{\gamma}}$, which for large $G$ and $\gamma >1$, may result
in a significantly milder hierarchy problem.

Of course, nailing down the pertinent values of $G$ and $\gamma$ is nothing short, in principle, of 
coarse-graining a putative and yet unknown theory of quantum gravity, not precisely a walk in the park...
 
However, this is precisely the context in which informed shortcuts "\`a-la Kolmogorov" might deliver 
clever insights into the problem, without facing its daunting mathematical complexity head-on \cite{LOLL}. 

The idea of adding "dissipative" terms to field-theoretical Lagrangians in order to
mitigate the hep hierarchy problem, if not resolve it altogether, might look outrageous at a first glance.
Not so fast, though: phenomenological Lagrangians are, by definition, coarse-grained versions of an allegedly
more fundamental microscopic theory, and dissipation is a well-known quintessential 
outcome of coarse-graining.

After all, this is much less exotic than extra-dimensions, warped spacetimes and other 
speculative scenarios which have been faring very high on the hep-theory agenda, notwithstanding 
their lack of experimental confirmation (so far).

\subsubsection{A wild numero-analogy}

Out of mere curiosity, we note that in order to bridge the $17$ decades between the Planck and the Higgs 
energy, $10^{19}$ versus $10^2$ Gev, via an inverse cascade with the same $h=1/3$ exponent as fluid turbulence, would take
a Reynolds number $Re \sim 10^{68}$, about a million times larger than the Reynolds number of the
Universe $Re_U \sim R_U/l_p$, where $R_U$ is the radius of the Universe and $l_p$ is the Planck scale.
Based on eq. (\ref{GAP_Cm}), a tiny $4$ percent increase, $h =17/45$, would fill exactly the bill!  

\section{Sneakiness and Fine-Tuning: catch me if you can!}

So far, we have discussed the role of nonlinearity as a scale bridger
and its ability to generate a broad bridge between the scale at which 
energy is injected at those at which it is dissipated.

This is what we call {\it Sneakiness}, important "things" tend to happen in very tiny 
corners of phase space, the "hide and seek" side of Nature, "catch me if you can"...

Note that "tiny" here is a relative statement: the coherent scale $l_c$ increases with increasing 
Reynolds number, but its ratio to the global scale decreases instead, which means that
in macroscopic units, the coherence scale gets progressively smaller, as per eq. (\ref{G3}).
   
Nonetheless, neglecting such sneaky regions, i.e. setting their size to zero, would result in total fiasco, 
since basically {\it all} of the fluid energy is dissipated in these thin boundary layers!
This is why the limit of zero viscosity, or, conversely, infinite Reynolds number, is a 
singular one, hence worth a million dollar!

Sneakiness results from nonlinearity, but not exclusively.

Another crucial trait of complex systems, hyper-dimensionality, does
just as well, in fact much more dramatically.

Consider the partition function of a fluid at equilibrium (a far cry from turbulence!):
$$
Z(T) = \sum_m e^{-H(q_m,p_m)/k_BT}
$$
 where $(q_m,p_m)$ denotes the $6N$ position-momentum coordinates of the 
 ensemble of $N$ molecules in the $m$-th microscopic realisation of a 
 macroscopic configuration with energy $E=H(p_m,q_m)$ at temperature $T$.
 
 At any given temperature $T$, the substantial contribution to the above 
 integral comes from a region of size $(k_BT)^{3N}$ in 6N-dimensional phase-space.
 
 For a fluid of just $N=100$ hard-spheres at the freezing temperature, Frenkel and Smit estimate that
 by sampling phase-space at random, only $1$ configuration in $10^{260}$  
 contributes significantly to the above integral \cite{DAAN}.  

We are really talking hide-and seek here!
 
Sneakiness connects with another major aspect of Naturalness, namely 
Fine-Tuning (FT) and Deep Cancellations (DC). 

As we have seen, a very pressing case of FT is the mass of the Higgs boson, around 100 Gev, which results
from cancellations of quantities at the Planck scale $10^{19}$ GeV, namely 17 digits!
This appears very weird and suspect, but is it really that weird?
After all,  DC's take place in other complex systems too, and sometimes 
with a true vengeance, as we shall show shortly. 

Before illustrating the point, a few technical details are in order.

The FT issues have been addressed by the so-called XY model \cite{XY}.
Here goes the story: let $z=x-y$ be an observable resulting from the tight "competition" between $x$ and $y$; in the XY literature this
is described as subtraction, because that's what it is indeed, a subtraction of renormalised contributions from the bare mass.
In the context of complex systems, however, $z$ is best viewed as a competition between two major contending
mechanisms, say Inertia and Dissipation for turbulence, or Energy and Entropy for thermodynamic systems. 
Indeed, free energy $F=E-TS$, is defined by the subtraction of "heat" from energy, but 
what it really measures in actual facts,  is the epic competition between energy and entropy in 
shaping up the material world around and within us.

Back to XY world.
The observable $z$ is said to be fine-tuned if it is much smaller than both 
its contributions $x$ and $y$, i.e.
$$
|z| \ll |x| \sim |y| 
$$
 If $x$ and $y$ are parameters of the theory, with no a priori estimate of their value, one 
 would naturally assume that they are equi-distributed in the unit square $[0,1] \times [0,1]$, i.e. $p(x,y)=1$.
 What is the probability of getting $|z| < \epsilon \ll 1$?
  
 A simple geometric argument helps: partition the unit square into $N=\epsilon^{-2}$ squarelets 
 of side $\epsilon$, the above probability is basically the number of squarelets 
 along the diagonal (sufficiently away from the origin), 
 versus the total number of squarelets, that is, approximately $\epsilon^{-1}/\epsilon^{-2}=\epsilon$.
 
 Hence, for small $\epsilon$, FT is a rare-event generator indeed.
 
 The assumption of a uniform pdf $p(x,y)=1$  is known as pro-naturalness position, as opposed 
 to the anti-naturalness position \cite{HOS}, which maintains the futility of using any 
 probability distribution at all, if the purpose is just to pick up a single value. 
 
 Yet, a uniform distribution is mostly natural if one assumes no correlation between 
 the two parameters, since such distribution maximises the entropy 
 $S[p]=-\int p(x,y) log p(x,y) dxdy = 0$, while any other $p(x,y)$ would give $S<0$.
 
On the other hand, suppose the parameters $x$ and $y$ are not independent, or suppose they
can be promoted to the status of dynamic variables with a low-scatter gaussian statistics, say
 $$
 p(x,y) \propto \epsilon^{-1} e^{-[(x-a)+(y-b)^2]/2 \epsilon^2}
 $$ 
 with $b=a(1\pm \epsilon)$, $\epsilon$ being the "temperature" of the thermal bath.
  
 In this case, Sneakiness would be fairly natural, no fine-tuning at all, similarly 
 to the partition function of the  aforementioned equilibrium fluid.
 
Of course, the case for sneakiness in statmech is incomparably more "natural", as it deals
with a large collection of genuinely dynamic variables, whose equations of motion can be
shown to sample the canonical distribution.
  
Why would/should (few) parameters in "theory space" , behave like Newtonian molecules?    

Far-fetching looms really large here.
  
Yet, on  a (very) loose basis, one might counter-argue that probability distributions supporting 
rare events are the rule and not the exception in complex systems.

Preferential attachment is a well-known mechanism in this respect: if you are a celebrity with 
many links to your site, the chance to get a new one is much larger than for the anonymous Joe Average.
In this respect, once value $x$ is realised, the probability of realizing a tight-close next
$y \sim x$ would be much higher than for a generic $y$ away from $x$.

To use a vivid metaphor I owe to Daan Frenkel \cite{DAAN2}, with the uniform distribution,
the idea of sticking your neck out of the hole of the first cannon ball, would make perfect sense. 
It is precisely the area argument we used to compute the probability $p(|z| < \epsilon)=\epsilon$.

Yet, in a Bayesian perspective, the strategy is utterly pointless: once you {\it know} 
that the ball has hit position $x=a$, if $x$ and $y$ are uncorrelated variables, the probability 
that the second ball hits the same location again, $y=a$, is the same as for 
any other value of $y$! 
It really helps nihil sticking your head out the first hole...    
And with preferential attachment, it is actually a literal recipe for suicide!
  
The upshot is that one should know more about the nature of parameter space before
the link between fine-tuning and Naturalness/sneakiness can be placed on a solid basis.        

This said, the question of whether Naturalness could somehow be facilitated by promoting
parameters to the status of dynamical variables, a not so infrequent practice in theoretical physics, 
see the invisible axion \cite{AXI}, neither in turbulence modelling \cite{SCI}, might perhaps
offer another route for cooperative explorations. 

\section{Mild and deep cancellations}

In this final section, we shall argue that deep cancellations hold in store interesting stories, 
regardless of whether they stand for genuinely physical ultra-near ties or just signal
inadequate representations instead. 
Either ways, profound lessons are likely to be lurking behind the corner. 

Let us begin by inspecting one of the most important "subtraction story" of all, namely the
very definition of free-energy $F=E-TS$, the literal currency of the material world.

Indeed, natural systems, and most notably soft matter and biological ones, live on the 
subtle Energy-Entropy edge, expressed by the free-energy minimum condition:
 $$
 \delta F = \delta E - T \delta S =0
 $$
also known as $f \equiv \frac{\delta E}{T \delta S} =1$, the champion number of naturalness.

The balance takes place up to fluctuations of order $1/\sqrt N$, with $N$ in the Avogadro's, hence
 twelve digits for just a very tiny cube centimetre of water. 
 
The argument is however bogus, because here the cancellation is expected to give {\it exactly}
zero, up to fluctuations, and not a very small, yet decidedly non-zero, number.

Yet, it is known that, unlike turbulence, soft matter and biology is a territory where energy and entropy compete
on a very thin rope, $F = E-TS \sim 0$, none of the two ever to win 
hands-down (if they did, we would not be here to witness).

More precisely, since free-energy is defined by its relation to the partition function: 
$$
Z=e^{-F/kT}
$$
it is clear that free-energy minima are the "important" states which contribute most 
to the partition function, the (overwhelming) rest of phase-space being practically 
inactive, as discussed above.  

So, let us inspect how close energy and entropy get for a specific and highly relevant 
biological process, protein folding. 

\subsection{Protein folding}

Proteins in the cell are manufactured in the ribosome in the form of linear chains 
of amino acids, the genetic sequence, also known as primary configuration.
Such linear sequence is however totally inane; in order to deliver its specific function, the 
protein must fold into a globular form, known as native state, the only one where it functions.

Failure to fold, or folding to the wrong state is no joke, as it spells major neurological 
diseases, Alzheimer's being just one grim case in point...  
In order to fold from the primary to the native state, the protein must manage to lower its free energy
content, which results from a highly complex competition between energetic send-down
and entropic send-up contributions \cite{FOLD}.

The energy budget scores approximately like this (in KJ/mole):
Hydrophobic forces (-200), Hydrogen-bonds  (-500), Van der Waals  (-50), Electrostatic (-50), for a total 
of  about $-800$. These are counter played by an entropy surplus of about +750,  due to chain configuration 
entropy (there is just one native state, as opposed to myriads of unfolded ones). 

The result is a final $\Delta F (unfold \to fold) = -50$, namely a rather 
modest, one-digit, cancellation: the competition is tight, but not unnaturally tight.

The Barbieri-Giudice FT functional \cite{BG} for this case reads  as $J=\frac{|\delta E|+T|\delta S|}{|\delta E-T \delta S|}$, delivering  
$J=1550/50 =31$, which is just a factor $3$ above the conventional threshold, hence not enough 
to qualify for FT status.

Thus, we conclude that, for all its breath-taking complexity, protein folding is, after all, 
"natural" enough, at least from the thermodynamic standpoint.

Less "natural", though, is the kinetics of protein folding, namely the way proteins manage to land 
on their native state in matter of minutes, navigating through monster-dimensional phase-spaces with
hundreds of dimensions, which would take {\it googols} of years (1 googol = $10^{100}$) to explore exhaustively.
Compare to the lowly $\sim 1.5 \times10^{10}$ years of the Universe, and what you get is
the famous puzzle known as Levinthal paradox, from MIT physicist Cyrus Levinthal \cite{LEV}. 

But this another flight \cite{WOL}.

\subsection{Many-body fermion ground states}

Next, let us consider a poignant example of deep cancellation, 
the ground state of many-body fermion systems \cite{TROY,CEP}.

Computing with fermions is notoriously painful, mostly on account of the very
"un-natural" {\it non-local} constraint represented by Pauli exclusion principle: the wave function of 
a quantum many-body function must change sign upon exchanging any two coordinates, say
$x_j$ and $x_k$:
$$
\Phi (x_1, \dots   {\bf x_j}  \dots   {\bf x_k}  \dots   x_N ) = - \Phi (x_1 \dots  {\bf x_k}  {\bf \dots x_j}   \dots  x_N )
$$ 
Bosons, on the other hand, are merrily indifferent to the exchange, hence their ground-state
wave function is symmetric (excited states are not, though).

A moment's thought shows that that Pauli's principle is pretty Un-natural, as it implies an 
instantaneous constraint between distant fermions, a loud scream against any 
principle of locality \cite{KAX}!

The consequences are infamously well known to practitioners: anti-commuting Grassmann numbers
instead of ordinary ones, non-positive definite probabilities, wildly oscillating integrals, being 
the typical plagues, vexing life of researchers in the field.

The fact that ground states of quantum many-body fermion systems are signed 
wave functions, implies that the ground state energy results from cancellations between the 
contributions of positive and negative components.

This has given rise to specific computational techniques which 
acknowledge the sign issue at the outset, the so called signed Quantum Monte Carlo method.

In this method, the wave-function splits into a non-negative and non-positive parts, respectively
 $$
 \phi^{\pm} = \frac{1}{2} (\phi \pm |\phi|) 
 $$
 At the ground state, the two contributions come to a tight balance, leading to
 a noise/signal ratio which diverges {\it exponentially} with the volume of the system 
 and its inverse temperature.
 The point is key, hence worth a few more details.
 
 Consider the expectation value of a given observable $A$ in the typical form
 \begin{equation}
 \label{A}
 <A>  = Z^{-1} \sum_i p_i A_i 
 \end{equation}
 where $\lbrace p_i \rbrace$ is a sequence of weights  and $A_i$
 is the value of the observable in state $i$.
 
 In the above $Z=\sum_i p_i$ is the partition function, which we assume to be finite and non-zero so that the 
 weights can always be normalised to unit, if so we wish.
 In a well-behaved scenario, all $p_i$'s are positive, and the above (hyper-dimensional) 
 integral can be performed by standard Monte Carlo.
 
 That's the happy boson story.
 
 But with fermions, because the wave function is signed, some $p_i$'s are necessarily negative, hence
 the expectation value $<A>$ comes from the difference between the positive and negative sets,
 $<A> = <A>_{+}  -  <A>_{-}$, a typical cancellation scenario.
 
 Is this cancellation deep? 
 Oh yes, pretty much so, in fact, exponentially deep!
 
 To see the point, let us write the probabilities in a way to expose their sign, i.e.
 $$
 p_i = |p_i| s_i
 $$
 where $s_i$ is the sign of $p_i$, i..e $s_i=+1$ if $p_i>0$, $s_i=-1$ if $p_i<0$, and $s_i=0$ if $p_i=0$.
 
 The expectation value eq. (\ref{A}), rewrites as follows
 \begin{equation}
 \label{A}
 <A>  = \frac{\sum_i |p_i| s_i  A_i}{\sum |p_i| s_i} 
 \end{equation}
 which can interpreted as the average of $As$ over the positive definite distribution $|p_i|$, namely
 $$
 <A>_p = \frac{<As>_{|p|}}{<s>_{|p|}}
 $$
 All is fine and well, since $|p|$ is now positive definite and all proceeds as in the merry boson scenario.
 
 Of course, this is too good to be true, the fermion problem must somehow
 raise its ugly head, and indeed it does so, through the statistics of the sign itself.
 In equations, 
 $$
 <s> = \frac{\sum_i s_i p_i}{\sum_i p_i}  
 = \frac{\sum_i |p|_i}{\sum_i p_i} 
 = \frac{Z}{Z_{|p|}} 
 $$  
 where we have used the identity $s_i^2=1$.

 By definition, the partition function relates to the free energy of the configuration, via 
 $Z=e^{-\beta N f}$,  where $N$ is the number of particles, $\beta=1/k_bT$ the inverse
 temperature and $f$ is the free energy density (per particle).
 
 Hence, we conclude that
 \begin{equation}
 \label{SIGN}
 <s> = e^{-\beta N (f-f_{|p|})}
 \end{equation}
 where $f$ is the free energy density of the fermion state and $f_{|p|}$ the one 
 of the corresponding boson system.
This quantity, which is in strict check of the energy of the fermion ground state, results
from an {\it exponential} cancellation between the free energy of 
fermion and boson degrees of freedom!

The analogy with FT is now pretty firm: $<s>$ is not, and {\it cannot}, be zero, if  it were 
the plus/minus symmetry would be unbroken and the ground the energy would be zero. 

It is however exponentially small at increasing $N$ and decreasing temperature.

Take a tiny $10^{-2} eV$ for $f-f_{|p|}$, at room temperature 
($kT=300$, i.e $\beta =40$ 1/ev) we obtain $e^{-0.4N}$, which 
with just $N=100$ fermions, gives $e^{-40} \sim  10^{-18}$, again 
precisely the gap between Planck to Higgs (and again, purely coincidental, of course!).

But fermions like to live in much colder environments, of the order of 1 Kelvin and even far below,
which boosts the exponent by a factor 300 and much more. 

It is fairly clear that the sign problem, all but a mathematical nicety, can be way more un-natural
than the Higgs mass and, depending on $N$ and $T$,  also readily in excess
of the 120 zero digits of the cosmological constant.

The reason may well be a very profound one, namely a fundamental flaw in using 
classical physics to model quantum systems, along the lines famously advocated by Feynman \cite{FEY}.

Indeed, on a quantum computer, the sign problem would dissolve by itself \cite{TROYER2}.

To the best of this author knowledge, the association of the sign problem with the
major fine-tuning problems in high-energy and cosmology is new. 
 Which does not make it useful by default; yet, given the tremendous amount of work developed
 by the condmat community to fight against the "astronomical" cancellations occurring
 in fermion quantum many-body problems, some form of cross-fertilisation does not sound that peregrine at all.
 
 This note was all about the prospects of importing ideas from non-hep scenarios to ease 
 the hep Naturalness problem. But the idea runs both ways: maybe SUSY ideas could spawn 
 new angles of attack to the tormented sign-problem. 
 After all, SUSY has already proved pretty valuable for the computation
 of low-lying eigenvalues of complex systems described by the 
 Fokker-Planck equation \cite{RISKEN}.
  
 Too good to be true? 
 Maybe, but ...  who knows?  

\section{Conclusions}

Summarising, we have discussed two major complex physical problems outside the 
high-energy realm, namely fluid turbulence and the ground state of quantum many-body fermions,
both of which show hints, if not distinct signatures, of  Un-naturalness.

For the case of turbulence, the key issue is Interscale Sensitivity and the
informing mechanism is nonlinearity, with the resulting scale-bridging cascades. 

We have also pointed out that in turbulence, the natural value $1$ (its own inverse) 
marks the border where two competing mechanisms, Inertia and Dissipation, come to a balance, a condition
which fixes the smallest scale at which coherence (field theory) manages to survive before
surrendering to microscopic disorganization.
In this respect, turbulence may not be "natural" at all, in that macroscopic physics, albeit
well separate from the microscopic world, is connected over a broad range of scales, hence
capable of "penetrating" pretty down close to the microscopic world.

The case of fermion ground states bears very sharply on the issue of Fine-Tuning. 
It is shown that as a consequence of the very "un-natural" character of Pauli's
exclusion principle, which flies patently in the face of locality, the energy of many-body
fermion ground states is controlled by an astronomically fine-tuned quantity, namely the
sign of the many-body wave function.   
This might be perfectly inline with Naturalness because, it is quite possible, in fact even likely based on
recent progress in the field, that fermions are un-natural only if one insists in simulating them on classical computers \cite{FEY}; on 
a quantum computer the sign problem would "melt like lemon drops", i.e. evaporate away by itself \cite{TROYER2}.

Both turbulence and fermion ground states show no special penchant for natural numbers,
meaning by this, one ($1$) and its close relatives, i.e. numbers which are largely
indifferent to inversion (self-dual).
Quite on the contrary, the fascination and subtlety of turbulence and fermions
rests upon their wild numbers, which cannot be inverted without pointing to 
completely different regimes. 

In other words, it is conjectured that the subtlety and fascination of these, and possibly many other complex 
physical systems, rests precisely on their "Un-naturalness", meaning by this 
the struggle to cope with Nature's tantalising attitude to play hide-seek in ultra-small
regions of phase space, through wild numbers, rare events, long-range, non-local correlations and deep cancellations. 

Whether standing for genuinely irreducible physics or artefacts due to inadequate mathematical representations,
the fascinating issues raised by the above Un-naturalness signatures are likely to offer stimulating
opportunities for cross-fertilisation between hep and the physics of complex systems. 

\section{Acknowledgments}

Illuminating discussions with Daan Frenkel, Gian Giudice, Sabine Hossenfelder, 
Tim Kaxiras, Giorgio Parisi, David Spergel and Mathias Troyer, are kindly acknowledged. 
Special thanks to Daan Frenkel for taking the pain of reading through the manuscript (and for 
declining paternity of the sticking-neck analogy in favour of \cite{DAAN2}...), to Gian Giudice
for critical reading and very valuable feedback, and to Mathias Troyer for sharing his highly informed 
insights on the sign problem.

The author also wishes to offer warm thanks to Fabiola Gianotti, for arranging a cameo-day
visit to CERN, where, in addition to being exposed to breath-taking experimental
facilities, he first happened to hear about Naturalness.

Financial support from the Center for Computational Astrophysics of the Simons Foundation and 
the European Research Council under the European Union Horizon 2020 Framework Programme 
(No. FP/2014-2020)/ERC Grant Agreement No. 739964 (COPMAT), are kindly acknowledged.


\begin{thebibliography}{99}
\bibitem{NAT}  We sometimes refer to Nature with expressions 
like "acts", "likes", "plays", as if she were equipped with an independent will. 
But this is of course only a metaphorical language.
\bibitem{LEV}
The back-of-the envelope calculation goes as follows. 
Take a $150$ amino acid chain, each amino acid being endowed with $10$ conformations, the total
number of configurations is $10^{150}$. With a conversion time scale of $10^{-12}$ s between conformations, the
time required by an exhaustive search is $10^{150-12}=10^{138}$ seconds, namely about $10^{130}$ years...
\end{thebibliography}
\end{document}